\documentclass[letterpaper]{article} 
\usepackage[draft]{aaai2026}  
\usepackage{threeparttable}
\usepackage{amsmath}
\usepackage{times}  
\usepackage{helvet}  
\usepackage{courier}  
\usepackage[hyphens]{url}  
\usepackage{graphicx} 
\usepackage{natbib}  
\usepackage{caption} 
\DeclareCaptionStyle{ruled}{labelfont=normalfont,labelsep=colon,strut=off} 
\usepackage{algorithm}
\usepackage{algorithmic}
\usepackage{booktabs}
\usepackage{xspace}

\newcommand{\pumpfun}{Pump.fun\xspace}

\usepackage{xcolor}
\newcommand{\rev}[1]{\textcolor{black}{#1}}

\title{Manufactured Authenticity: How Behavioral Signals Are Produced and Misread in Online Markets}

\author{
    Hou-Wan (Darren) Long\textsuperscript{\rm 1},
    Yun Huang\textsuperscript{\rm 2}\thanks{Yun Huang is the corresponding author.}
}
\affiliations{
    \textsuperscript{\rm 1}Tepper School of Business, Carnegie Mellon University\\
    \textsuperscript{\rm 2}University of Illinois Urbana-Champaign\\
    darrenlong@cmu.edu, yunhuang@illinois.edu
}

\begin{document}

\maketitle

\begin{abstract}
Online platforms increasingly ask users to infer credibility from behavioral signals whose production conditions are opaque. We introduce \emph{manufactured authenticity}: the platform-mediated production of observable activity that interfaces invite users to read as organic participation. Using 996,834 \rev{cryptocurrency} tokens and 50,895 creator wallets from Pump.fun\rev{, a Solana-based token launchpad}, we examine how legitimacy signals are produced in an automation-enabled online market. We identify four production strategies (high-velocity launching, passive launching, template-content creation, and organic-style creation) that generate distinct but interface-similar cues of activity, narrative plausibility, and participation. High-velocity and template-content strategies emerge as alternatives rather than cumulative layers: the former produces activity at scale, while the latter produces descriptions associated with higher reply engagement. Yet visible engagement does not imply broader diffusion: template-like content attracts replies without greater ownership dispersion, and \rev{tokens associated with non-organic production strategies exhibit more concentrated investor holdings than organic-style tokens}. We argue that platform governance should move beyond detecting manipulation toward making signal provenance legible. We propose \emph{authenticity disclosure} as a design approach for surfacing production conditions without requiring definitive claims about automation, AI authorship, or actor intent.
\end{abstract}

\section{Introduction}

What counts as credible participation when the visible traces of activity can be cheaply generated, coordinated, and scaled?\rev{\footnote{Throughout this paper, ``token'' refers to cryptocurrency tokens---digital assets launched on blockchain platforms---not to tokens in the natural language processing sense.}} Across online platforms, users routinely infer legitimacy from behavioral signals such as engagement counts, apparent participation diversity, temporal activity, market momentum, and visible community formation. Yet as automation tools, coordinated account networks, AI-assisted operators, and autonomous agents become embedded in digital platforms, these signals no longer reliably indicate independent human interest or organic collective behavior~\cite{gorwa2020bot,hancock2020aimc}. They may instead be produced through repeatable, machine-mediated workflows.

This creates a credibility and governance problem. Platform interfaces often present behavioral traces as evidence of social reality while revealing little about how those traces were produced. Users are therefore asked to make consequential judgments from signals whose provenance is opaque. Prior work shows that people struggle to distinguish automated participation from human activity, and that the heuristics used to make such judgments are exploitable~\cite{jakesch2023heuristics}. Inflated engagement metrics can increase susceptibility to low-credibility information~\cite{avram2020metrics}, while social bots can amplify low-credibility content early in information cascades~\cite{shao2018lowcredibility}. The problem is therefore not only whether users can detect automation, but whether platform interfaces transform machine-producible behavior into persuasive evidence of popularity, momentum, trust, or legitimacy.

We conceptualize this condition as \emph{manufactured authenticity}: the platform-mediated production of observable activity that interfaces invite users to read as organic participation. This framing shifts attention from isolated bad actors to the production and presentation of social evidence. In automation-enabled markets, authenticity is not simply a property of an actor, account, or transaction; it is an inference made from platform-visible traces. When those traces can be generated cheaply and repeatedly, authenticity becomes a design and governance problem: platforms must decide whether to expose, obscure, or monetize the production conditions behind the signals users rely on.

\begin{figure*}[t]
    \centering
    \includegraphics[width=0.95\textwidth]{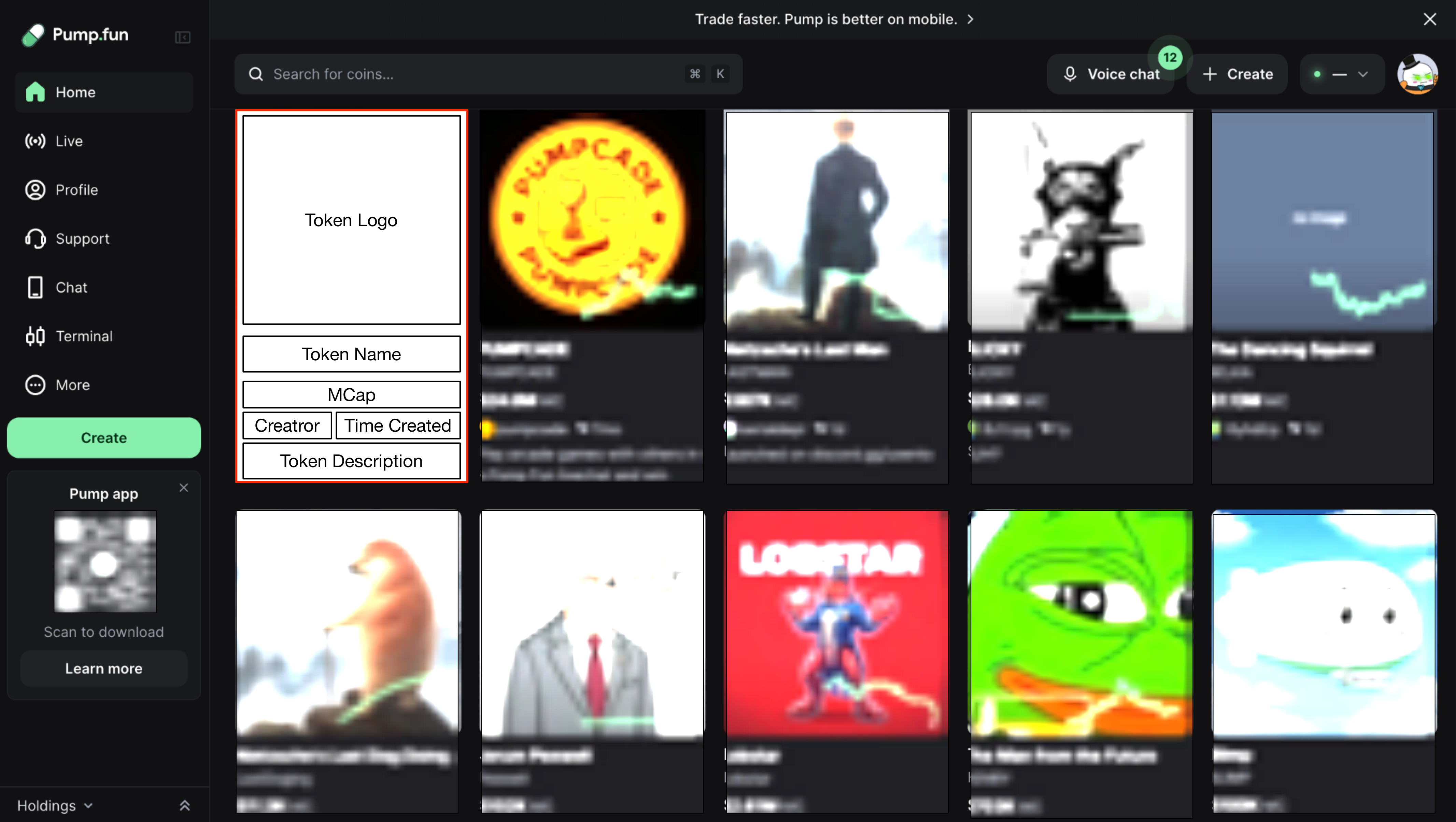}
    \caption{Annotated token card on the \pumpfun discovery page.}
    \label{fig:interface-card}
\end{figure*}

We study manufactured authenticity through \pumpfun, Solana's dominant token launchpad \rev{(Figure~\ref{fig:interface-card})}. We use \pumpfun not because memecoins are exceptional, but because the platform provides a high-resolution setting for a broader problem: users must make rapid, financially consequential judgments from interface-visible behavioral traces whose origins are opaque. Participants evaluate tokens through cues such as early buyer activity, transaction frequency, apparent wallet diversity, price movement, replies, creator identity, and perceived market momentum. These cues make tokens legible as opportunities, while also inviting users to treat visible activity as evidence of organic community formation.

Existing research on crypto manipulation has primarily examined token-level abuse, including pump-and-dump schemes, rug pulls, wash trading, bot activity, and retail loss~\cite{hamrick2019pdump,xu2019pumpdump,gandal2018price,wu2025rphunter,li2025hnn4rp,mongardini2025midsummer,mancino2025memecoinphenomenon,niedermayer2024bots,gan2024ammwash}. This work explains how individual assets are manipulated, but less attention has been paid to how legitimacy signals are repeatedly produced across assets by concentrated production infrastructures. For platform governance, the object of analysis should therefore not be only the manipulated token, but the production system that makes activity appear socially meaningful.

\rev{We use \emph{legitimacy} to mean a token's perceived appropriateness or credibility as a participation opportunity, and \emph{authenticity} to mean the correspondence between observed signals and the independent, organic processes they are taken to represent. \emph{Signal provenance}, our analytical focus, concerns who or what produced a trace, through what process, and at what scale.}

We ask: \emph{how is authentic participation manufactured across multiple production layers, and how do these production strategies relate to token-level attention, success, and ownership diffusion?} We address this question through three research questions:

\begin{itemize}
\item \textbf{RQ1: How concentrated is token production across creator wallets and behavioral production strategies?}

\item \textbf{RQ2: What behavioral, content, and market fingerprints distinguish these production strategies?}

\item \textbf{RQ3: How do automation-consistent production, template-scored content, and creator velocity relate to engagement, graduation, and holder concentration?}
\end{itemize}

\rev{To answer these questions, we analyze 996,834 tokens and 50,895 creator wallets using on-chain activity and platform metadata from January 2024 to April 2026. Creator-level features and K-means clustering identify four production strategies: High-Velocity Launchers, Passive Launchers, Template-Content Creators, and Organic-Style Creators. High launch velocity and frequent template use characterize distinct profiles. Template-like content is associated with higher reply engagement without greater ownership dispersion, while tokens associated with non-organic production strategies exhibit more concentrated investor holdings.}

\rev{This paper makes three contributions. First, we introduce \emph{manufactured authenticity} as a framework for examining how repeatable production supplies behavioral cues that may be interpreted as evidence of legitimacy. Second, we provide a large-scale account of concentrated and behaviorally distinct token production on \pumpfun.\footnote{\url{https://anonymous.4open.science/r/Manufactured-Authenticity-2410}} Third, we show that surface-level engagement and ownership dispersion can diverge, motivating \emph{authenticity disclosure}: interface cues that expose the production conditions behind visible activity.}

\section{Related Work and Background}
\label{sec:related-work}

This section positions manufactured authenticity in relation to coordinated inauthentic behavior, automated signal production, and behavioral detection of automation. We then introduce \pumpfun as a setting where these issues become empirically visible: users evaluate assets through platform-visible signals whose production histories are largely hidden.

\subsection{From Coordinated Actors to Produced Signals}

\rev{Work on \emph{coordinated inauthentic behavior} (CIB) examines accounts acting together to amplify content~\cite{gleicher2017cib,starbird2019disinfoops}, while \emph{astroturfing} describes organized attempts to simulate grassroots support~\cite{ratkiewicz2011astroturf,metaxas2012astroturf}. These approaches foreground participating actors and networks; manufactured authenticity adds attention to the recurring production and interface presentation of behavioral traces.}

\rev{Crypto-market research documents pump-and-dump schemes, wash trading, rug pulls, bot-driven trading, liquidity manipulation, and retail investor loss~\cite{hamrick2019pdump,xu2019pumpdump,gandal2018price,wu2025rphunter,li2025hnn4rp,mongardini2025midsummer,mancino2025memecoinphenomenon,niedermayer2024bots,gan2024ammwash}. Studies of Telegram and Twitter show how coordinated buy signals can inflate token prices before organized sell-offs~\cite{kamps2018moonlambo,xu2019pumpdump}. Information cascade theory~\cite{bikhchandani1992cascades} offers a related mechanism: observable signals can encourage participation even when their origins are opaque. These studies motivate examining recurring production across assets alongside individual manipulation events.}

\rev{The dynamics we study also parallel content creation incentives on non-crypto platforms. On YouTube~\cite{bishop2019algorithmiccreator} and TikTok~\cite{klug2021tiktok}, platform algorithms reward visible engagement, creating incentives for creators to produce signals of popularity regardless of whether those signals reflect authentic audience interest. \pumpfun operates under a similar logic: interface-visible metrics such as reply counts, market capitalization, and apparent momentum make tokens legible as opportunities, even when the underlying participation structure is narrow.}

\subsection{Behavioral Evidence of Automation}

A parallel literature studies how automated or script-assisted activity can be inferred from behavior. Social bot research shows that automated accounts often leave temporal, linguistic, and network signatures~\cite{gorwa2020bot,ferrara2016rise,varol2017online}. Interaction timing, content regularity, and network structure can distinguish automated from human-operated accounts even when individual messages appear plausible~\cite{cresci2017paradigm}.

\rev{Recent work on AI-agent platforms extends this approach. \citet{li2026moltbook} study Moltbook, a platform populated by large language model agents, using the coefficient of variation of inter-post intervals to distinguish autonomous from human-directed activity. Low-variation posting suggests scheduled operation, whereas irregular timing is more consistent with human prompting. This approach motivates our use of temporal regularity as evidence of \emph{automation-consistent behavior}, alongside measures of formulaic description text.}

\begin{figure*}[t]
  \centering
  \includegraphics[width=\textwidth]{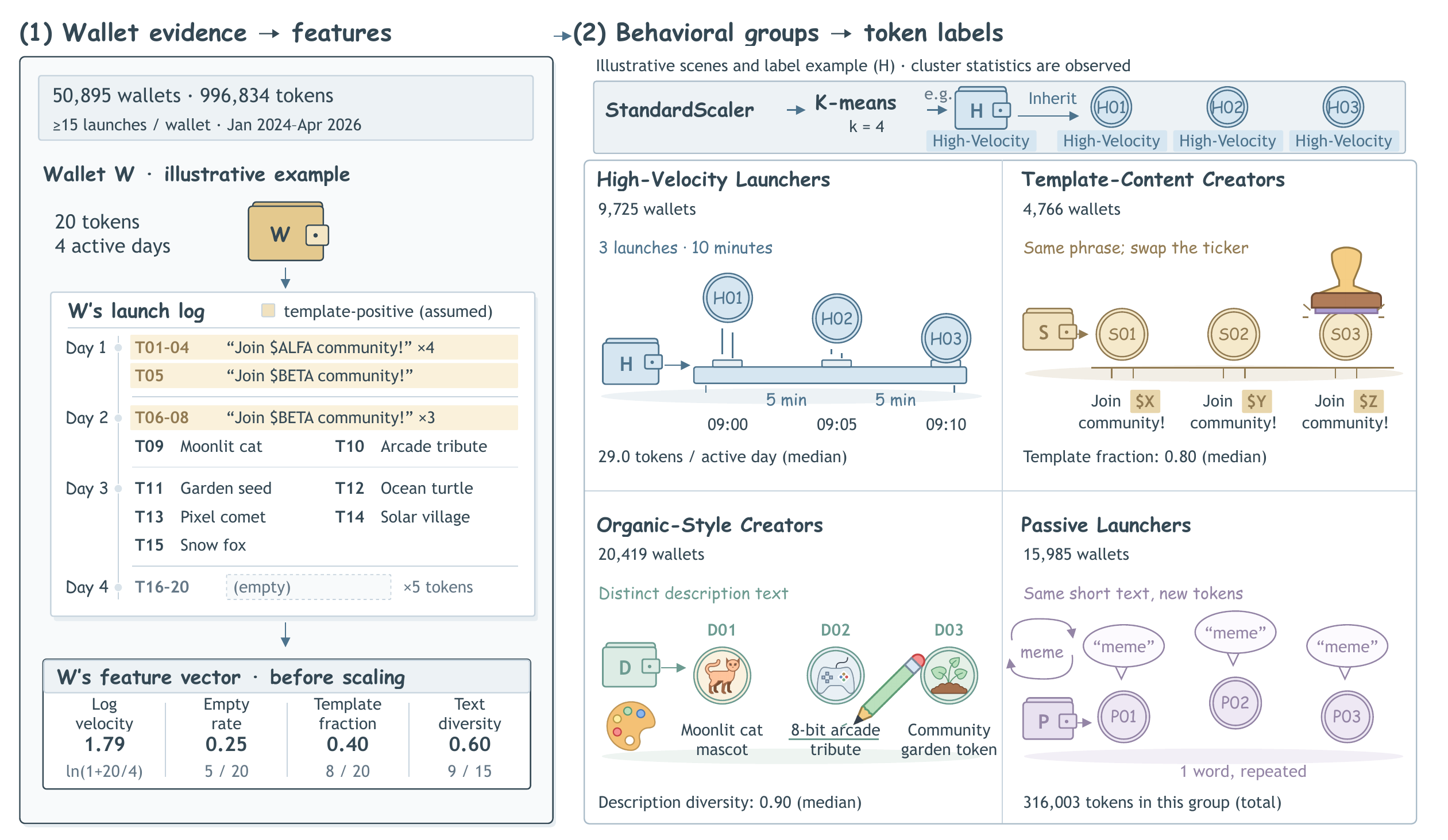}
  \caption{\rev{Wallet-level feature construction, behavioral clustering, and token-label inheritance. The launch log is illustrative.}}
  \label{fig:pipeline}
\end{figure*}

\subsection{\pumpfun as an Automation-Enabled Market}

\rev{\pumpfun provides an automation-enabled setting in which scripts, bots, AI-agent tools, and human-in-the-loop workflows can support token creation, trading, promotion, and protocol interaction~\cite{li2026moltbook,gorwa2020bot}.}

\pumpfun is a useful case because any wallet can create a tradeable token in seconds, without identity verification or cooldowns and with a low launch fee. By mid-2025, the platform had hosted over 11.9 million token launches~\cite{chainalysis2025manipulation,marino2026pumpfun}, while 98.6\% of tokens never reached the bonding-curve graduation threshold required for public exchange listing~\cite{soliduslabs2025,marino2026pumpfun}. In the absence of audited fundamentals, stable issuer identities, or verified community histories, users infer legitimacy from visible cues such as early buying activity, price movement, transaction frequency, token comments, wallet dispersion, and apparent community momentum.

The surrounding ecosystem makes this production logic explicit. Public tools include trading bots that monitor token creation and execute rapid buy/sell decisions,\footnote{Chainstack's pump.fun trading bot implementation: \url{https://github.com/chainstacklabs/pumpfun-bonkfun-bot}.} services that coordinate wallet activity and engagement-like signals,\footnote{Pump Fun Bots services: \url{https://smithii.io/en/pump-fun-bots/}.} AI-agent frameworks for Solana protocols,\footnote{Solana AI-agent tooling: \url{https://solana.com/skills}.} and generative systems for token names, slogans, images, and promotional narratives.\footnote{Meme Coin Generator GPTs: \url{https://app.aiprm.com/gpts/g-6741b795bed08191935933b4134ea2ed/meme-coin-generator}.} These tools need not be fully autonomous to matter. Even partial automation increases the speed, scale, and repeatability of participation.

\rev{Trading bot documentation (e.g., Chainstack's \pumpfun bot, BonkBot, Photon) describes reply counts, holder counts, and market capitalization as inputs to automated buy-signal filters. Related work shows that engagement counts can increase susceptibility to low-credibility information~\cite{avram2020metrics} and that prior ratings can induce herding~\cite{muchnik2013social}.}

The significance of this automation depends on interface design. As shown in Figure~\ref{fig:interface-card}, \pumpfun compresses market activity into token cards that display a logo, name, market capitalization, creator identifier, creation time, and short description. These cues make newly launched tokens legible as comparable opportunities. Yet the same interface does not show whether a creator has launched hundreds of prior tokens, whether a wallet is short-lived, or whether early activity is concentrated across related wallets. This creates a gap between signal visibility and signal provenance: activity produced by independent users, coordinated wallets, scripts, or mixed human-machine workflows can appear similar at the point of decision-making.

\section{Data and Methods}
\label{sec:methods}

This study analyzes token creation behavior on \pumpfun from January 2024 through April 2026. Because on-chain data identify wallets rather than persons, organizations, or automated agents, creator identities are analyzed at the wallet level. Our empirical strategy therefore focuses on observable production behavior: how creator wallets launch tokens, how they use description text, and how these production patterns are associated with token-level outcomes.

We first construct creator-level features that capture four dimensions of production behavior: launch velocity, empty-description rate, promotional-template content, and description diversity. We then use K-means clustering to group creator wallets into four behavioral clusters. These clusters summarize distinct production profiles across the creator population and are used descriptively throughout the analysis.

\rev{Figure~\ref{fig:pipeline} illustrates creator-level feature construction, behavioral clustering, and the assignment of creator cluster labels to tokens. The example launch log is constructed, and highlighted descriptions are assumed template-positive. A \textbf{creator wallet} launched at least one token during the observation period; a \textbf{buyer wallet} purchased at least one. \textbf{Operational lifespan} is the number of calendar days between a creator wallet's first and last observed token launch. The four behavioral features and cluster labels are defined below.}\looseness=-1

\subsection{Data Collection}
\label{sec:data-collection}

We combine creator-wallet records, token metadata, holder distributions, and wallet lifecycle data.

\textbf{Creator data.}
We identify creator wallets using Dune Analytics and order them by the total number of tokens created in descending order. \rev{We retain wallets with at least 15 launches to construct behavioral features from repeated observations, yielding 50,895 creator wallets. The threshold applies to creator launch histories; our metadata corpus covers a subset of these wallets' launches. Section~\ref{sec:rq1} reports the sample's population coverage.}

\textbf{Token metadata.}
Using the creator set obtained from Dune, we collect token-level metadata from the public \pumpfun API. These records include token name, ticker, description text, image URI, creator wallet, market capitalization, community reply count, and bonding-curve completion status (\texttt{is\_complete}). We supplement these records with on-chain graduation data from Dune Analytics. After deduplication, the analysis dataset contains 996,834 tokens.

\textbf{Holder data.}
To measure early ownership concentration, we collect holder information for a stratified sample of 1,000 tokens per behavioral cluster, for a total of 4,000 tokens. Top-20 holder distributions are retrieved from Solana mainnet RPC using \texttt{getTokenLargestAccounts} and \texttt{getTokenSupply} via the Helius endpoint. \rev{Because \texttt{getTokenLargestAccounts} returns all token accounts including protocol-controlled accounts---the bonding-curve contract that holds unsold supply and, for graduated tokens, the Raydium AMM liquidity pool---we exclude any account holding more than 50\% of total supply before computing concentration metrics. HHI is then computed from the renormalized shares of the remaining non-protocol holders. Tokens with no remaining non-protocol holders (i.e., all supply is held by the bonding curve) receive HHI~$= 10{,}000$, indicating zero investor diffusion.} For each token, we compute the Herfindahl--Hirschman Index (HHI), scaled from 0 to 10,000, using the percentage shares of the \rev{non-protocol} top-20 holders.

\textbf{Wallet lifecycle and launch timing data.}
For all 50,895 creator wallets, we retrieve transaction history using \texttt{getSignaturesForAddress} RPC calls through Helius, with up to 100 signatures per wallet. From these signatures, we derive three wallet lifecycle variables: \texttt{wallet\_age\_days}, defined as the span between the oldest and newest observed on-chain signature; \texttt{is\_likely\_fresh}, indicating whether the observed wallet age is less than 14 days; and \texttt{had\_prior\_activity}, indicating whether any observed signature predates the wallet's first \pumpfun token launch.

We also compute launch timing measures directly from each token's \texttt{created\_timestamp}. These include median inter-launch interval, launch-hour Shannon entropy, weekend-launch fraction, and single-day peak launch count. Wallet lifecycle and timing variables are used only as post-hoc behavioral descriptors and are not included as clustering inputs.

\subsection{Behavioral Feature Construction}

We characterize each creator wallet using four creator-level behavioral features.

\textbf{Launch velocity.}
Launch velocity measures how quickly a creator wallet produces tokens during its observed period of activity. For each creator, we compute the number of tokens launched per active day and use the transformed variable
\[
\texttt{log\_vel} = \log(1 + \text{tokens per active day}).
\]
This transformation reduces the influence of extreme high-volume creators while preserving the ordering of launch intensity.

\textbf{Empty-description rate.}
\rev{For each creator wallet, \texttt{empty\_rate} is the proportion of token launches with no description text.}

\textbf{Promotional-Template Score.}
For each token description, we compute a Promotional-Template Score, denoted \texttt{desc\_template\_score} $\in [0,1]$. This score measures the presence of formulaic promotional language in token descriptions. It combines three sub-signals: (i) opener patterns, such as ``Introducing\ldots'' or ``Welcome to the future of\ldots''; (ii) body patterns, such as ``decentralized ecosystem'', ``seamless experience'', or ``unprecedented opportunity''; and (iii) lexical diversity, measured using type-token ratio, where lower diversity is consistent with more templated text.

\rev{Empty descriptions and descriptions identified as spam receive a score of zero. The score measures lexical patterns rather than authorship: human writers, snippet libraries, and generative tools can produce similar patterns.}

\rev{\textbf{Threshold usage.} The creator-level clustering input \texttt{frac\_template} is the proportion of a creator's token records with score~$>0$. Token-level regressions use the continuous score, standardized to zero mean and unit variance.}

\textbf{Description diversity.}
\rev{For each creator, description diversity (\texttt{desc\_diversity}) is the number of distinct non-empty description texts divided by the number of non-empty token descriptions.}

\subsection{Behavioral Cluster Classification}

We classify the 50,895 creator wallets into four behavioral clusters using K-means clustering. The clustering inputs are all measured at the creator level: log launch velocity (\texttt{log\_vel}), empty-description rate (\texttt{empty\_rate}), promotional-template fraction (\texttt{frac\_template}), and description diversity (\texttt{desc\_diversity}). Features are standardized using \texttt{StandardScaler}. We estimate K-means with $k=4$, seed 42, and \texttt{n\_init=20}.

\rev{\textbf{Choice of $k$ and method.} We select $k=4$ based on the silhouette score (0.388), Calinski--Harabasz index (35,588), and interpretive coherence. While $k=2$ yields the highest silhouette score (0.432), it collapses the four-way production strategy distinction central to our analysis. The log-transform of launch velocity reduces right-skew from 52.7 (raw) to 0.93, and StandardScaler normalization further mitigates scale effects, making K-means appropriate despite the original distributional skew. We also evaluated Gaussian Mixture Models and hierarchical Ward clustering at $k=4$, both producing high agreement with the K-means partition (ARI $> 0.85$).}

Cluster labels are assigned after model estimation based on each cluster's median feature profile. The cluster with the highest median launch velocity is labeled \emph{High-Velocity Launchers}. Among the remaining clusters, the one with the highest median promotional-template fraction is labeled \emph{Template-Content Creators}. Of the two remaining clusters, the one with the lowest empty-description rate and highest description diversity is labeled \emph{Organic-Style Creators}; the final cluster is labeled \emph{Passive Launchers}.

\rev{These clusters summarize observed wallet behavior rather than verified actor identities or automation status. The ``organic-style'' label is comparative: the group has the highest description diversity (0.90), the lowest empty-description rate (0.05), and a median template-content fraction of 0.25, the lowest among groups with template usage.} Each token inherits the behavioral cluster label of its creator wallet.

\rev{Outcome regressions retain continuous predictors, including empty-description rate and Promotional-Template Score.}

To assess cluster stability, we conduct a bootstrap analysis across 50 random 80\% subsamples. The resulting partitions are highly reproducible, with a mean adjusted Rand index (ARI) of 0.990, standard deviation of 0.006, and minimum value of 0.980. \rev{\textbf{Sensitivity to $k$.} At $k=3$ (ARI $= 0.679$ vs.\ $k=4$), Passive Launchers and High-Velocity Launchers merge, losing the distinction between high-speed and low-effort production; the Template-Content and Organic-Style clusters are preserved. At $k=5$ (ARI $= 0.708$), the High-Velocity cluster splits into two velocity tiers (61\%/38\% of the original), while Template-Content (84\%), Organic-Style (84\%), and Passive (86\%) clusters remain largely intact. The paper's qualitative conclusions---concentrated production, multiple distinct strategies, and the speed-vs-content tradeoff---hold for both $k=3$ and $k=5$.}

\subsection{Post-hoc Behavioral Descriptors}

After clustering, we use wallet lifecycle and launch timing measures to describe how the resulting groups differ operationally. These measures are not used to form the clusters.

Wallet lifecycle descriptors include \texttt{wallet\_age\_days}, \texttt{is\_likely\_fresh}, and \texttt{had\_prior\_activity}. These variables help distinguish wallets with short observed on-chain histories from wallets with broader prior activity. Launch timing descriptors include median inter-launch interval, launch-hour Shannon entropy, weekend-launch fraction, and single-day peak launch count. Together, these measures provide additional evidence on whether clusters differ in temporal regularity, launch concentration, and wallet history. \rev{These post-hoc descriptors are used exclusively to characterize cluster profiles in Table~\ref{tab:quadrant-summary}; they are not included as predictors in the regression analyses of Section~\ref{sec:rq3}.}

\rev{Table~\ref{tab:quadrant-summary} reports median launch velocity (tokens per active day) and template-content fraction, alongside post-hoc medians for observed wallet age (days), inter-launch interval (seconds), and launch-hour entropy. Entropy is measured in bits, with a maximum of $\log_2 24 \approx 4.58$. Reply and graduation rates are the proportions of tokens receiving at least one community reply and completing the bonding curve, respectively. Fresh denotes the proportion of wallets with observed age below 14 days; Wknd denotes the proportion of launches on weekends.}

\section{Results}

\subsection{RQ1: Token Supply Is Produced by Concentrated Creator Regimes}
\label{sec:rq1}

We first examine whether the stream of new tokens on \pumpfun reflects broad-based creator participation or concentrated production. To users, the platform appears as a continuous flow of new opportunities. Behind that flow, however, token supply is highly uneven: a small upper tail of creator wallets produces a disproportionate share of all launches.

\rev{During the observation period, \pumpfun hosted approximately 4.32~million creator wallets, with a median of 1~launch and a mean of 2.57; 87.6\% launched exactly one token. Our behavioral-analysis sample comprises 50,895 wallets with $\geq$15 launches, or 1.2\% of all creators. We collected metadata for 996,834 tokens from these wallets, equivalent to approximately 9.0\% of platform launches. This is the collected metadata corpus, not these creators' total launch output. At the platform level, the top 0.1\% of wallets (approximately 4,320) account for 26.6\% of all launches.}

\begin{figure}[t]
\centering
\includegraphics[width=\columnwidth]{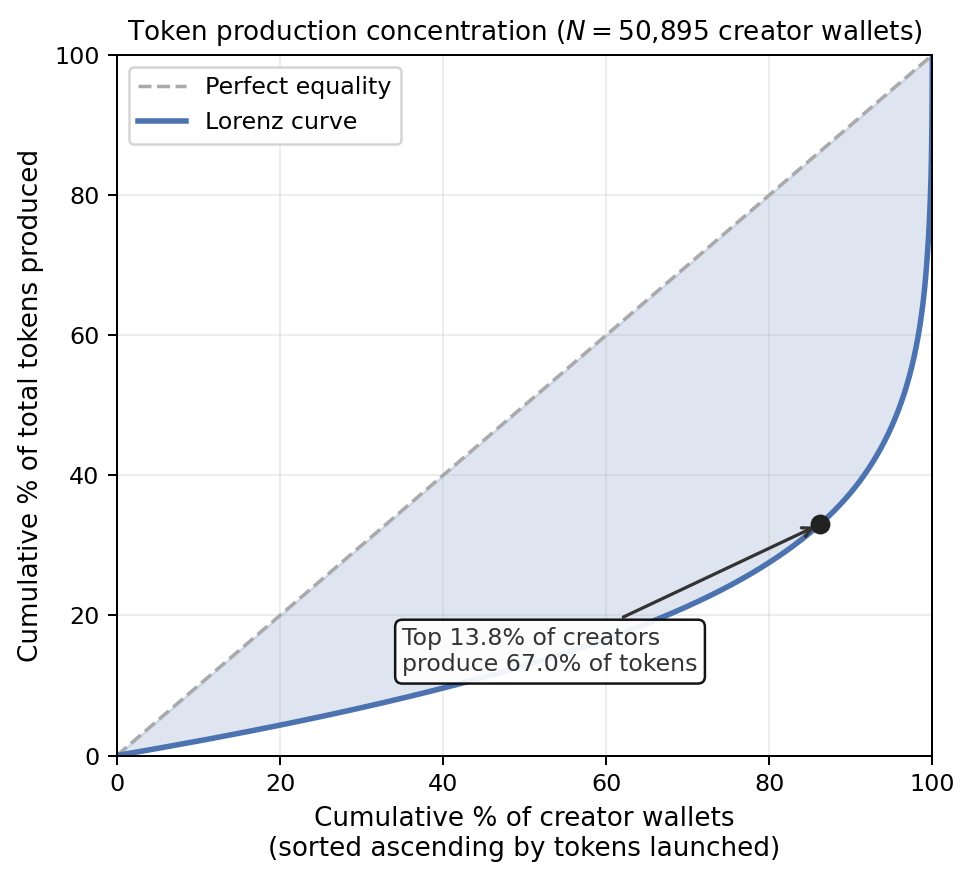}
\caption{\rev{Lorenz curve of token production within the $\geq$15-launch creator sample.}}
\label{fig:creator-concentration}
\end{figure}

\rev{Figure~\ref{fig:creator-concentration} shows the Lorenz curve within the $\geq$15-launch creator sample: the top 13.8\% of these wallets account for 67.0\% of their token production. The curve excludes approximately 4.27~million other wallets and describes concentration within the selected creator sample. Activity-based filtering can affect inequality estimates~\cite{deltas2003gini,warrens2018gini}, so these results should be distinguished from platform-wide concentration.}

Wallet lifecycle patterns reinforce the interpretation of token supply as production infrastructure. Among High-Velocity Launcher wallets, 55\% have an observed on-chain age below 14 days, with a median age of 7.1 days. Organic-Style Creators show the opposite pattern: a median wallet age of 62.2 days and only 28\% fresh wallets. The median gap between a wallet's oldest observed signature and its first \pumpfun launch is under one day for High-Velocity Launchers, compared with over 40 days for Organic-Style Creators. These patterns suggest that high-output supply is supported not only by rapid launch workflows, but also by a renewable stock of short-lived creator identities.

Together, these findings show that \pumpfun's visible stream of token opportunities is not organic-looking by default. It is structured by concentrated production infrastructure that repeatedly generates the activity cues users encounter at the interface. The next section examines the distinct strategies within this production infrastructure.

\subsection{RQ2: Multiple Strategies Manufacture Legitimacy}

We next examine how concentrated token supply is organized into distinct forms of legitimacy-signal production. The clusters are not simply volume categories. As Table~\ref{tab:quadrant-summary} shows, they correspond to different ways of making tokens appear active, persistent, narratively plausible, or comparatively organic.

High-Velocity Launchers manufacture legitimacy through volume and momentum. They account for 187,421 tokens across 9,725 wallets and have the highest median launch velocity, at 29.0 tokens per active day. Their median inter-launch interval is only 309 seconds, their launch-hour entropy is the lowest of all clusters (1.97 bits), and they record zero weekend launches. This combination of speed and temporal regularity suggests compressed production sessions that leave market-visible traces at high frequency.

Passive Launchers manufacture legitimacy through persistence and platform density. They account for 316,003 tokens across 15,985 wallets, making them the second-largest token-producing cluster. Their median launch velocity is much lower than that of High-Velocity Launchers (4.2 tokens per active day), and their median inter-launch interval is much longer (4,561 seconds), but they still produce repeated low-information launches at scale. Their reply rate is the lowest across clusters (60.4\%), consistent with a strategy that contributes to the background volume of the platform rather than to token-specific narrative appeal.

Template-Content Creators manufacture legitimacy through narrative plausibility. They account for fewer tokens (91,006) and wallets (4,766), but have the highest median template-content fraction (0.80). They are not the fastest launchers: their median velocity is 11.5 tokens per active day, well below that of High-Velocity Launchers (29.0). Yet their tokens receive the highest reply rate across clusters (71.6\%), suggesting that formulaic descriptive content is associated with greater surface-level interaction.

Organic-Style Creators provide a comparatively diverse baseline. They are the largest wallet group, with 20,419 wallets, and produce the largest number of tokens, 402,404. Their median launch velocity is the lowest (3.3 tokens per active day), their median inter-launch interval is the longest (6,509 seconds), and their wallet age is the highest (62.2 days). They also show relatively high reply engagement (69.7\%) and the highest weekend-launch fraction (21\%), consistent with slower and more temporally dispersed production.

The contrast between High-Velocity Launchers and Template-Content Creators is central. High-Velocity Launchers have the highest median velocity (29.0) but a median template-content fraction of 0.00, whereas Template-Content Creators have the highest template-content fraction (0.80) but substantially lower velocity (11.5). This shows that high-velocity production and template-content production do not collapse into a single ``sophisticated bot'' strategy. Activity at scale and narrative plausibility are empirically distinct pathways for producing legitimacy signals. One possible explanation is an operational tradeoff: rapid-launch strategies may minimize textual complexity to preserve speed, while template-content strategies may invest more in narrative plausibility.

Community response follows the content gradient more than the velocity gradient. Template-Content Creators and Organic-Style Creators have the highest reply rates, at 71.6\% and 69.7\%, respectively. High-Velocity Launchers receive lower reply engagement (64.9\%), while Passive Launchers receive the weakest engagement (60.4\%). This pattern suggests that descriptive substance, whether comparatively organic or formulaic, helps elicit interaction more than launch speed alone. Market graduation follows a different pattern. Graduation remains rare across all clusters, but High-Velocity Launchers have the highest graduation rate (3.24\%), followed by Passive Launchers (2.29\%), Organic-Style Creators (1.46\%), and Template-Content Creators (1.10\%). Thus, template-content use is associated with higher replies but not higher graduation. Market success appears more closely aligned with operational intensity or repeated experimentation than with narrative plausibility alone.

Overall, RQ2 shows that manufactured authenticity is not a single tactic. It is produced through multiple operational logics: volume, persistence, narrative plausibility, and comparatively diverse participation. These strategies shape which tokens become visible and interaction-worthy, but they do not map cleanly onto market success.

\begin{table*}[t]
\centering
\caption{Behavioral cluster profiles: production and outcomes (left); wallet history and launch timing (right).}
\label{tab:quadrant-summary}
\footnotesize
\resizebox{\textwidth}{!}{%
\begin{tabular}{l r r r r r r | r r r r r}
\toprule
 & \multicolumn{6}{c|}{Production \& Outcomes} & \multicolumn{5}{c}{Wallet Lifecycle \& Launch Timing} \\
\cmidrule(r){2-7}\cmidrule(l){8-12}
Cluster & Tokens & Wallets & Med.~vel & Frac.~template & Reply rate & Grad.~rate & Age~(d) & Fresh & Intvl~(s) & Hr-ent & Wknd \\
\midrule
High-Velocity Launchers   & 187,421 &  9,725 & 29.0 & 0.00 & 64.9\% & 3.24\% &  7.1 & 55\% &    309 & 1.97 &  0\% \\
Passive Launchers    & 316,003 & 15,985 &  4.2 & 0.00 & 60.4\% & 2.29\% & 41.3 & 34\% &  4,561 & 2.94 & 20\% \\
Template-Content Creators  &  91,006 &  4,766 & 11.5 & 0.80 & 71.6\% & 1.10\% & 22.3 & 46\% &    574 & 2.36 & 10\% \\
Organic-Style Creators & 402,404 & 20,419 &  3.3 & 0.25 & 69.7\% & 1.46\% & 62.2 & 28\% &  6,509 & 2.98 & 21\% \\
\bottomrule
\end{tabular}
}
\end{table*}

\subsection{RQ3: Visible Engagement Diverges from Diffusion}
\label{sec:rq3}

\rev{RQ3 asks whether visible attention is accompanied by broader market participation. We relate replies, market capitalization, and graduation to creator production signals, then compare ownership concentration across clusters. Table~\ref{tab:regression} reports OLS estimates for 996,834 tokens, with standard errors clustered across 50,895 creator wallets.}

\rev{Tokens with more template-like descriptions receive more replies in the main specification, as do tokens from faster-launching creators. Engagement is lower when creators leave more descriptions empty. These features explain little of the variation in replies ($R^2 = 0.009$): they describe average associations, but leave substantial differences between individual tokens unexplained.}

\rev{The reply patterns do not carry over uniformly to market outcomes. Associations with market capitalization are close to zero, and the velocity estimate is not distinguishable from zero. Graduation remains rare, occurring for approximately 2\% of tokens. Empty-description rate and velocity have the largest positive associations with graduation, while template score has a smaller association. For scale, each predictor's average marginal effect in the logistic model is below one percentage point per standard deviation.}

\rev{Accounting for launch month, description length, and image presence absorbs the template-score association with replies, while the directions for velocity and empty-description rate remain. Reply-model fit rises to $R^2 = 0.372$. The template-content pattern therefore depends on the comparison context. Estimates for graduation and replies are similar when restricted to tokens at least 30 or 90 days old.\footnote{The adjusted model uses 28 launch months (January 2024--April 2026). Logistic graduation ORs (95\% CIs; creator-wallet-clustered SEs) are 1.40 $[1.36, 1.44]$ for empty-description rate, 1.24 $[1.20, 1.28]$ for velocity, and 1.06 $[1.04, 1.08]$ for template score (all $p < 0.001$). Corresponding average marginal effects are $+0.66$, $+0.42$, and $+0.12$ percentage points per standard deviation. The velocity estimate for log(mcap) has $p = 0.19$. The 30-day graduation estimate for empty-description rate is $\beta = +0.0067$ $[+0.0061, +0.0073]$, compared with $+0.0066$ in the full sample.} Full robustness results are available in the anonymous code repository.}

\rev{
\begin{table}[t]
\centering
\caption{OLS estimates with standardized predictors; brackets show creator-wallet-clustered 95\% confidence intervals.}
\label{tab:regression}
\footnotesize
\setlength{\tabcolsep}{2pt}
\renewcommand{\arraystretch}{1.05}
\begin{tabular*}{\columnwidth}{@{\extracolsep{\fill}}l c c c@{}}
\toprule
Predictor & log(replies) & log(mcap) & is\_complete \\
\midrule
Template        & $+0.033$             & $+0.001$             & $+0.001$             \\
score           & {\scriptsize$[+0.028, +0.039]$} & {\scriptsize$[+0.000, +0.001]$} & {\scriptsize$[+0.001, +0.001]$} \\[2pt]
Empty-desc.     & $-0.109$             & $-0.004$             & $+0.007$             \\
rate            & {\scriptsize$[-0.117, -0.100]$} & {\scriptsize$[-0.004, -0.003]$} & {\scriptsize$[+0.006, +0.007]$} \\[2pt]
log(velocity)   & $+0.016$             & $+0.000$             & $+0.005$             \\
                & {\scriptsize$[+0.007, +0.024]$} & {\scriptsize$[-0.000, +0.001]$} & {\scriptsize$[+0.004, +0.006]$} \\
\midrule
$R^2$           & 0.009                & 0.001                & 0.004                \\
\bottomrule
\end{tabular*}
\end{table}
}

\subsubsection{Early holder concentration.}

\rev{As described in Section~\ref{sec:data-collection}, we exclude protocol-controlled accounts (bonding-curve and AMM liquidity pools holding $>$50\% of supply) before computing HHI, ensuring that concentration reflects investor ownership rather than unsold protocol-held supply.}

\begin{figure*}[t]
  \centering
  \includegraphics[width=\textwidth]{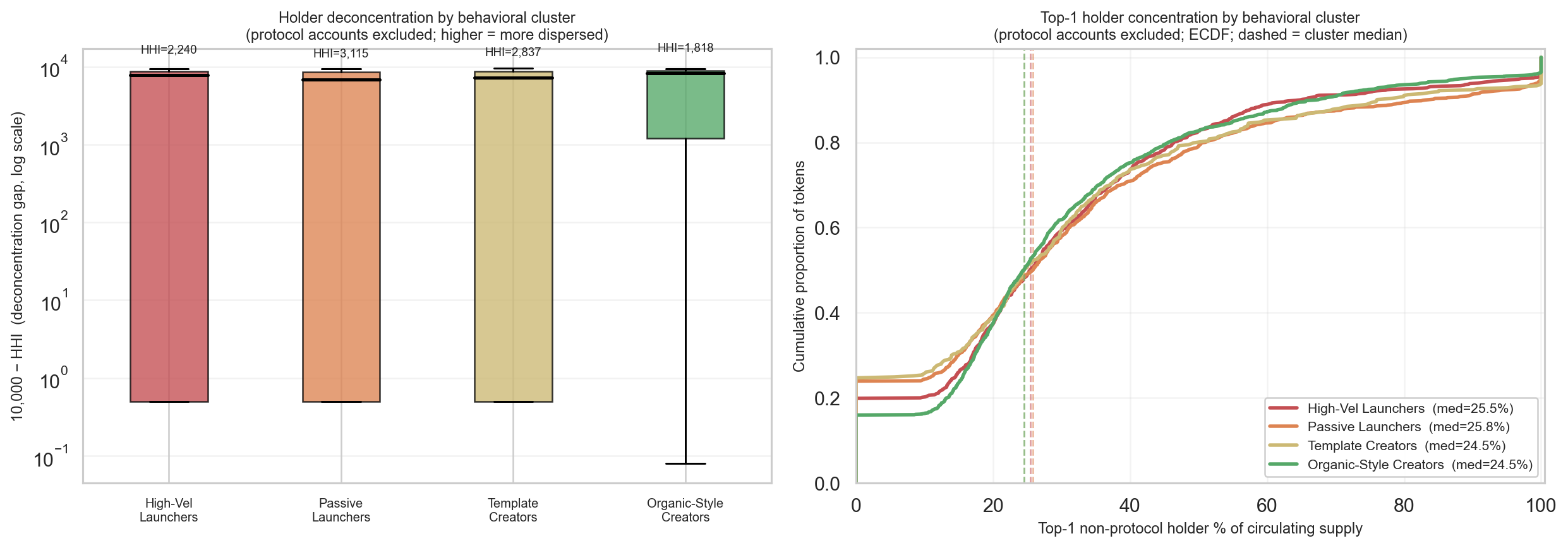}
  \caption{\rev{Holder concentration by cluster, excluding protocol accounts: deconcentration gap ($10{,}000-\mathrm{HHI}$; left) and ECDF of the top-1 holder share (right).}}
  \label{fig:hhi}
\end{figure*}

Figure~\ref{fig:hhi} shows that visible engagement does not imply broader ownership diffusion. \rev{After excluding protocol accounts, holder concentration varies substantially across clusters (Kruskal-Wallis $H = 71.36$, $p < 0.001$). Organic-Style Creator tokens have the most dispersed ownership (median HHI $=1{,}818$), followed by High-Velocity Launchers and Template-Content Creators; Passive Launcher tokens are the most concentrated (median HHI $=3{,}115$).}

The left panel reports the deconcentration gap, defined as $10{,}000-\mathrm{HHI}$, where $\mathrm{HHI} = \sum_i s_i^2$ over percentage ownership shares \rev{among non-protocol holders} and $\mathrm{HHI}=10{,}000$ indicates single-holder dominance. Larger values indicate more dispersed ownership. Organic-Style Creator tokens have the largest deconcentration gap\rev{, while Passive Launcher tokens have the smallest}. The right panel shows the \rev{same ordering using the top-1 non-protocol holder share}.

This is the key divergence. Template-Content Creators generate the highest reply rate in Table~\ref{tab:quadrant-summary}, but their tokens do not show the same ownership dispersion as Organic-Style Creator tokens. \rev{Passive Launchers and High-Velocity Launchers generate activity at scale, but their tokens remain more concentrated among non-protocol holders.} Thus, the interface-visible signs of activity, narrative plausibility, and engagement do not necessarily correspond to broader participation. Manufactured tokens can appear active or legible at the surface while remaining concentrated in their early ownership structure.

\section{Discussion}

\subsection{Manufactured Authenticity as Signal Infrastructure}

\rev{Our results identify concentrated token supply, distinct creator production profiles, and a separation between reply engagement and investor-level ownership dispersion on \pumpfun. These findings connect visible token activity to the production histories of creator wallets.}

\rev{The theoretical implication is that visible engagement and participation breadth should be treated as separate dimensions. Manufactured authenticity captures how repeatable production can supply visible activity even when investor ownership remains narrow. This motivates examining a signal's production history alongside its magnitude.}

\rev{The relevant unit of interpretation extends beyond the individual token. A token card presents one launch; a creator's portfolio shows how it fits into sustained production, low-information releases, or recurring promotional language. Reading these levels together connects a visible object to the process that generated it. Similar-looking activity can therefore reflect distinct production histories, with different relationships to replies, graduation, and ownership.}

\subsection{Platform Design and the Production of Authenticity}

\rev{Three design conditions help interpret these results. \textbf{Weak identity persistence} allows repeated launching without durable reputation across tokens. \textbf{Low-friction creation} lowers the cost of experimentation, scripting, and high-volume production. \textbf{Opaque signal provenance} leaves creator launch histories and ownership concentration outside the token card. This gap between visible metrics and production context connects our findings to research on deceptive affordances in interface design~\cite{gray2018darkpatterns,mathur2019darkpatterns}.}

\rev{The adjusted reply model motivates closer attention to launch context. The template-score association is absorbed after adding launch-month fixed effects, description length, and image presence, while velocity and empty-description rate retain their direction. Comparisons within launch cohorts and within the same wallets over time could help separate changing production practices from changes in the creator population.}

\rev{Platform evaluation could similarly separate activity volume from participation breadth. Alongside reply counts, an audit could report how many distinct creator wallets appear in discovery results and how often users encounter launches from the same wallet. It could also track the holder concentration of tokens gaining visibility. These measures would connect production patterns to the allocation of attention while distinguishing engagement, creator representation, and ownership dispersion.}

\rev{The visibility--diffusion gap also motivates questions for other platforms organized around engagement metrics. Information cascade theory~\cite{bikhchandani1992cascades} offers one account of how observable signals can encourage participation, while research on YouTube and TikTok documents incentives for producing visible engagement~\cite{bishop2019algorithmiccreator,klug2021tiktok}. Future work could test provenance disclosures in settings such as product reviews, crowdfunding, and content recommendation.}

\subsection{Making Signal Provenance Legible}
\label{sec:disclosure}

\rev{We propose \emph{authenticity disclosure}: interface cues that show the observable production conditions behind visible activity. Creator velocity, observed wallet history, repeated descriptions, and holder concentration provide information that platforms could present alongside engagement metrics.}

\rev{Four disclosure types follow from our findings. A \textbf{sustained-velocity disclosure} could show launch volume over a specified period, for example: ``This creator has launched 47 tokens in the past 5 days.'' A \textbf{disposable-creator disclosure} could show high launch volume alongside a short observed on-chain history; in our data, 55\% of High-Velocity Launcher wallets have an observed on-chain age under 14 days. A \textbf{template-content disclosure} could show repeated promotional phrasing. A \textbf{concentration disclosure} could report how strongly early ownership is concentrated among holders.}

\rev{A disclosure could pair a compact summary with inspectable evidence. Launch counts should name their time window; repetition labels should show the comparison set and example descriptions; concentration labels should identify the holder snapshot, holder coverage, and excluded protocol accounts. Readers could follow each summary to the underlying launches or text matches. Consistent definitions and observation windows would support comparisons across tokens and make changes in the summary traceable to the observations.}

\rev{An experiment could compare identical token cards displaying engagement metrics alone, metrics plus creator history, or metrics plus history and holder concentration. Holding descriptions and engagement counts fixed would isolate the contribution of provenance cues. Outcomes could include perceived popularity, inferred breadth of participation, and confidence in those judgments. A direct test would ask whether high reply counts still imply widely distributed ownership to users when a concentration cue is visible. Comprehension and willingness to engage should be measured separately.}

\rev{Authenticity disclosure also creates tradeoffs. Actors may adapt to disclosure thresholds, users may over-trust unflagged tokens, and velocity or freshness cues may stigmatize legitimate newcomers. Disclosures should therefore be layered and revisable, show the underlying observations, and be framed as provenance cues rather than fraud labels.}

\subsection{Limitations and Ethical Considerations}
\label{sec:limitations}

\rev{We analyze wallet-level associations using observational data. These patterns do not establish causality or actor intent, and graduation models lack censoring adjustment. We report aggregate results without deanonymizing wallet controllers; user interpretations remain a question for future experiments.}

\section{Conclusion}

\rev{Across 996,834 tokens and 50,895 creator wallets on \pumpfun, we identify four distinct production strategies. High launch velocity and frequent template use characterize alternative profiles: the former produces activity at scale, while the latter is associated with higher reply engagement. These results support manufactured authenticity as a framework for examining how repeated production supplies visible market activity.}

\rev{The central finding is that visibility and diffusion diverge. Template-content tokens receive more replies without corresponding ownership dispersion, while high-velocity and passive tokens remain more concentrated among non-protocol holders than organic-style tokens. These associations motivate \emph{authenticity disclosure}: showing creator velocity, observed wallet history, content templating, and holder concentration alongside engagement metrics.}

\clearpage
\bibliography{references}

\end{document}